\documentclass[article]{IEEEtran}
\ifCLASSINFOpdf
\usepackage{graphics,epsf,epsfig,fullpage,rotate,subfigure,lineno,amsmath,amssymb}
\else
\fi

\begin{document}

\title{\textsc{MaGe} - a {\sc Geant4}-based Monte Carlo Application Framework for Low-background 
Germanium Experiments}

\author{Melissa Boswell$^{1}$, 
        Yuen-Dat Chan$^{2}$,
         Jason A. Detwiler$^{2,3}$,
         Padraic Finnerty$^{4,13}$,
	Reyco Henning$^{4,13}$, 
	Victor M. Gehman$^{1,3}$
	Rob A. Johnson$^{3}$, 
	David V. Jordan$^{5}$, 
	Kareem Kazkaz$^{3,6}$, 
	Markus Knapp$^{7}$, 
	Kevin Kr\"oninger$^{8,9}$, 
	Daniel Lenz$^{8}$,
	Lance Leviner$^{12}$,
	Jing Liu$^{8}$, 
	Xiang Liu$^{8}$, 
	Sean MacMullin$^{4,13}$,
	Michael G. Marino$^{3}$, 
	Akbar Mokhtarani$^{2}$, 
	Luciano Pandola$^{10}$, 
	Alexis G. Schubert$^{3}$, 
	Jens Schubert$^{8}$,
	Claudia Tomei$^{10,11}$,
	Oleksandr Volynets$^{8}$

\thanks{
  $^{1}$~Los Alamos National Laboratory, Los Alamos, NM, USA,
  $^{2}$~Lawrence Berkeley National Laboratory, Berkeley, CA, USA, 
  $^{3}$~University of Washington, Seattle, WA, USA, 
  $^{4}$~University of North Carolina, Chapel Hill, NC, USA, 
  $^{5}$~Pacific Northwest National Laboratory, Richland, WA, USA,	
  $^{6}$~Present address: Lawrence Livermore National Laboratory, Livermore, CA, USA, 
  $^{7}$~Physikalisches Institut, Universit\"at T\"ubingen, Germany, 
  $^{8}$~Max-Planck-Institut f\"ur Physik, Munich, Germany,  
  $^{9}$~Present address: II. Institute of Physics, University of G\"ottingen, Germany,
  $^{10}$~INFN, Laboratori Nazionali del Gran Sasso, Assergi, Italy,
  $^{11}$~Present Address: INFN, Sezione di Roma, Rome, Italy,
  $^{12}$~North Carolina State University, Raleigh, NC, USA,
  $^{13}$~Triangle Universities Nuclear Laboratory, Durham, NC, USA
        }
	
} 

\markboth{Transactions on Nuclear Science (TNS)}%
{Boswell \MakeLowercase{\textit{et al.}}: {\sc MaGe} - a GEANT4 based
Monte Carlo framework for low-background experiments}

\maketitle

% Commands added during rewrite by RH, 10/17/08

\newcommand{\Gerda}{\textsc{Gerda}}
\newcommand{\GeSS}{$^{76}\mathrm{Ge}$}
\newcommand{\GF}{\textsc{Geant4}}
\newcommand{\MaGe}{\textsc{MaGe}}
\newcommand{\MJ}{\textsc{Majorana}}
\newcommand{\nubb}{$0\nu\beta\beta$}

% Use \GF and not \G4 to keep LaTex from barfing.

\begin{abstract}
We describe a physics simulation software framework, \MaGe, that is based on the
\GF\ simulation toolkit. \MaGe\ is used to simulate the response of  
ultra-low radioactive background radiation detectors to ionizing radiation, specifically
the \MJ\ and \Gerda\ neutrinoless double-beta decay 
experiments. \MJ\ and \Gerda\ use high-purity germanium detectors to search for the neutrinoless 
double-beta decay of \GeSS , and \MaGe\ is jointly developed between these two collaborations. 
The \MaGe\ framework contains the geometry models of common objects, prototypes,  
test stands, and the actual experiments. 
It also implements customized event generators, \GF\ physics lists, and output 
formats. All of these features are available as class libraries that are typically compiled
into a single executable. The user selects the particular experimental setup 
implementation at run-time via macros. 
The combination of all these common classes into one framework 
reduces duplication of efforts, eases comparison between simulated data and experiment, and simplifies the addition of new detectors to be simulated. This paper focuses on the software framework, custom event generators, and physics lists. 
\end{abstract}

\begin{IEEEkeywords}
Monte Carlo, neutrinoless double-beta decay, Germanium detectors, 
Geant4, radiation detection, low background.
\end{IEEEkeywords}

% \fi
\IEEEpeerreviewmaketitle

% INTRODUCTION

\section{Introduction}
\label{se:Introduction}

\MaGe\ (MAjorana-GErda) is a \GF-based~\cite{Agostinelli:2002hh,Allison:2006} physics simulation 
software 
framework jointly developed by the \MJ\ and
\Gerda\ collaborations~\cite{Aalseth:2004yt, Schoenert:2005}. Both experiments will
search for the neutrinoless double-beta decay (\nubb\ decay)
of the \GeSS\ isotope using arrays of isotopically enriched High-Purity Germanium (HPGe) detectors.  
The discovery of \nubb\ decay is the only
practical way to determine if the neutrino is a Majorana particle. For
further details on the physics motivation, see the review article
in~\cite{Avignone:2007fu}. The current lower limit on the \nubb\ decay  
half-life of \GeSS\ is 
$1.9\times10^{25}\,\mathrm{years}$~\cite{Klap:2001}, making this decay extremely rare if it exists. 
This requires great care to reduce experimental backgrounds from naturally occurring radioactivity 
and cosmic rays. This is achieved via careful material selection and assay, a deep underground 
location, passive and active shielding, and analysis cuts.
The purpose of \MaGe\ is to simulate the response of the \MJ\ and \Gerda\ 
experiments to ionizing radiation from backgrounds, calibration sources, and \nubb\ decays. 
\MaGe\ is also used to simulate the response of prototype detectors, test-stands, and low-background 
assay systems.  
In the prototyping phase, the simulation
is used as a virtual test stand to guide detector design, to estimate the
effectiveness of proposed background reduction techniques, and to estimate
the experimental sensitivity. During experimental operation, \MaGe\ will be used to simulate and characterize unexpected
backgrounds and determine the ultimate sensitivity of the experiments. It will also provide 
probability distributions for signal extraction analyses.
The combination of the two collaboration's simulation package into one framework 
reduces duplication of efforts, eases comparison between simulated data and experiment, and simplifies the addition of new simulated detector geometries. \MJ\ and \Gerda\ are currently constructing detectors with tens of kilograms of enriched isotopes, but it is the goal of parts of the two collaborations to merge and pursue a joint effort towards a tonne-scale germanium experiment. Having a joint simulation package during the very early phases will ease this future integration. This paper focusses on the software framework, custom event generators, and physics lists of \MaGe .

The code and physics requirements are given in section~\ref{section:Requirements}. The code structure of \MaGe\ is discussed in section~\ref{section:structure}. 
Section~\ref{section:physics} describes the implemented \GF\ physics lists that are optimized for low-energy (sub-keV to few MeV), low background applications. Validation of \MaGe\ 
simulations against experimental data is discussed in Sect.~\ref{section:validation}.
Conclusions are provided in the last section.

% ---------------------------------------------------------------------
% Requirements
% ---------------------------------------------------------------------

\section{Requirements}
\label{section:Requirements}

The requirements for the \MaGe\ framework can be subdivided into physics requirements and software requirements. 
The physics requirements define the physical processes that have to be simulated to find the response of the detectors. 
Software requirements are driven by the use of the \GF\ toolkit as basis for \MaGe\ and the anticipated end-users. 

\subsection{Physics Requirements} 
\label{se:physics_requirements}

\GF\ is a simulation toolkit that uses Monte Carlo techniques to simulate the propagation of particles and nuclei through matter. It has extensive capabilities to simulate different experimental geometries, propagating particles, and particle interactions, and has the foundation of the physics requirements for \MaGe. 
The choice of \GF\ over other packages as the basis for \MaGe\ was motivated
by its flexibility and active development within the particle and
medical physics communities, as well as its C++ and object-oriented structure. \GF\ is open-source and allows collaborations with members from multiple countries to use it. It is a standard simulation tool for LHC experiments and will likely be supported for at least another decade. 

The most stringent
requirements for \MaGe\ are the proper simulation of the relevant
background sources for \nubb\ decay
experiments. \GF\ fulfills some of these requirements, specifically it simulates:
\begin{enumerate}
\item Electromagnetic interactions of electrons and $\gamma$-rays at MeV and keV energies.
\item Radioactive isotope decay chains and nuclear de-excitations. 
\item Interactions of thermal and fast neutrons. 
\item Electromagnetic and hadronic showers
initiated by cosmic-ray muons.
\item Penetration depths and ionization energy loss profiles of 
$\alpha$-particles. 
\end{enumerate}
The list of implemented physics models in \MaGe\ was optimized for
low-background, underground physics
applications~\cite{Bauer:2004an}, with an emphasis on low-energy
interactions and hadronic interactions resulting from cosmic-ray
spallation. 
\MaGe\ has implemented different selections of models in its \GF\ physics list. These are tailored to fit specific physics applications, and the required processes can be selected by the user to optimize computation time and data storage requirements. For example, the simulation of muon propagation through rocks requires different models than that of background from radioactive decays in detector components. Many
important tuning parameters, such as the	 production cuts for $\delta$-rays and soft bremsstrahlung photons, may also be set by the user. 
Specific details of the physics lists implemented in \MaGe\ are discussed in Section~\ref{section:physics}. The \MaGe\ code is regularly updated and ported in order to make
it compatible with the most recent \GF\ releases.\\

Other physics processes that are not part of \GF\ had to be simulated. These include:
\begin{enumerate}
\item Electric-field solvers are required to simulate the trajectories of charge carriers inside the HPGe crystals under the influence of the biasing field.
\item Generators of electronic waveforms by charge carriers in the HPGe detectors as they drift inside the crystal towards the collection electrodes. %Analysis of this pulse-shapes is a key method for reduction of background and is described below.
\item Electronic transfer functions of generated pulses into simulated detector pulses. 
\item A generic surface sampler that uniformly and randomly samples points on arbitrary surfaces had to be implemented to simulate surface alpha contaminations. The algorithm of this sampler is described in Ref.~\cite{surfacesampler}.
\item Event generators for physics processes other than normal nuclear decay, such as two neutrino double-beta decay for different models, \nubb\ decay, and cosmic-ray muons at depth.  
\end{enumerate}
Some of the existing aspects of \GF\ had to be extended or improved to fulfill the requirements of \MaGe . These are described later in this paper.

\subsection{Software Requirements}
\label{se:software_requirement}

The use of \GF\ as the basis of \MaGe\ made C++ the natural choice for \MaGe , and \MaGe\ makes full use of the object-oriented nature of C++ and \GF{}. The software framework is required to:
\begin{enumerate}
\item Allow run-time selection of detector configuration, event generators and output format. 
\item Allow parallel and independent
development of different branches of the code. For example, the development of the geometric descriptions of the \Gerda\ and \MJ\ detector geometries should proceed independently and with minimal interference. 
\item Ease-of-maintenance over the full life time of the
experiments.
\item Perform simulations with different
configurations by physicists that are inexperienced in programming.   
\end{enumerate}
The collaboration does not maintain the \MaGe\ source code for public 
release, since most of components are experiment-specific and not useful beyond the two collaborations. However, some select components, such as physics lists and event generators, are provided to 
interested parties on a case-by-case basis. The object-oriented nature of the \MaGe\ framework makes the transfer of 
such code to other users straightforward. 

% ---------------------------------------------------------------------
% structure of mage 
% ---------------------------------------------------------------------

\section{Structure of \MaGe\ }
\label{section:structure}

To realize the requirements outlined in the previous section we subdivided a simulation task into different  components. These components are geometries, generators, output formats, and physics lists. The user instantiates one instance of a class corresponding to each component at run-time, typically via a \GF\ messenger in a macro file. A component may then also instantiate other components, and have its own macro commands that allow the user to further refine the simulation parameters. This design allows the simulation of many different detectors, prototypes and
validations to be performed within the same executable using
the same physics processes, geometries and other codes. This greatly eases cross-comparisons and reduces coding and debugging effort. The components are described in this section. 

\subsection{Geometries}
\label{se:geometries}
\begin{figure}[tbh]
\flushleft
\includegraphics[width=0.4\textwidth]{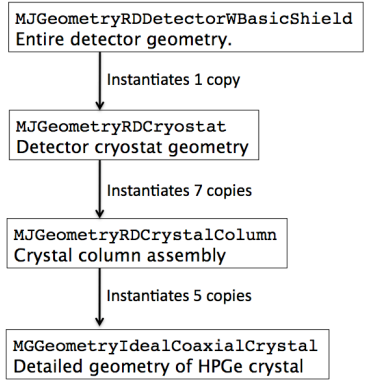}
\caption{Classes used to create the geometry of the \MJ\ {\sc Demonstrator}. This detector consists of 35 crystals arranged in 7 columns of 5 crystals each. The crystals are placed inside a cryostat that in turn is placed inside a layered shield. This diagram illustrates how this design is realized in the simulation. Each of these classes can be instantiated on its own, or as part of a larger geometry, such as this. A rendering of this geometry is shown in Fig.~\ref{fig:MJDEMOrendering}.}\label{fig:MJDEMOGeometry} 
\end{figure}

\begin{figure}[tbh]
\centering
\includegraphics[width=0.45\textwidth]{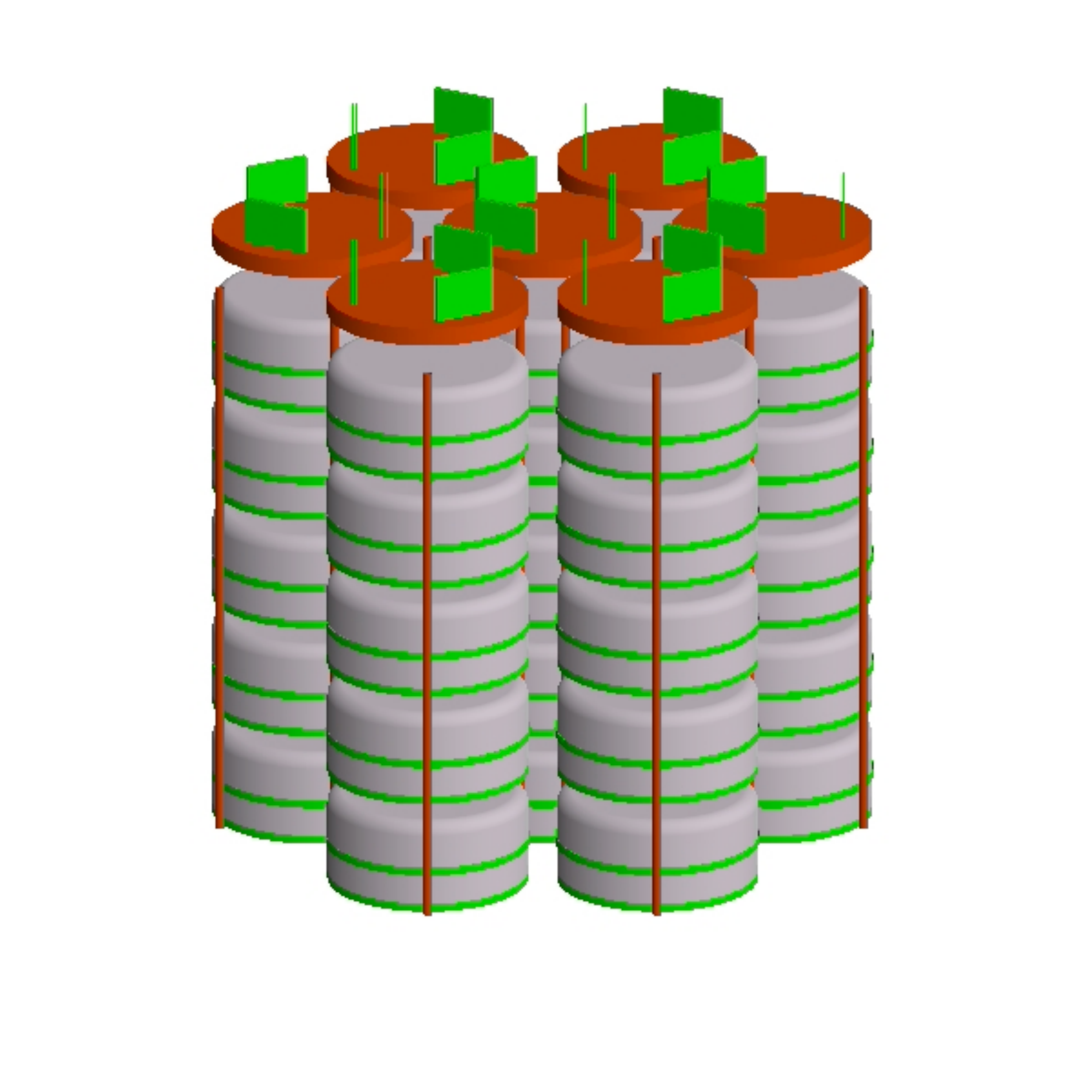}
\caption{Rendering of \MJ\ {\sc Demonstrator} with cryostat and shields removed that was created using the class structure in Fig.~\ref{fig:MJDEMOGeometry}. Each of the 7 crystal columns with associated support structures is a stand-alone class, as is each of the 5 crystals in a column. }
\label{fig:MJDEMOrendering}
\end{figure}

\begin{figure}[tbh]
\centering
\includegraphics[width=0.45\textwidth]{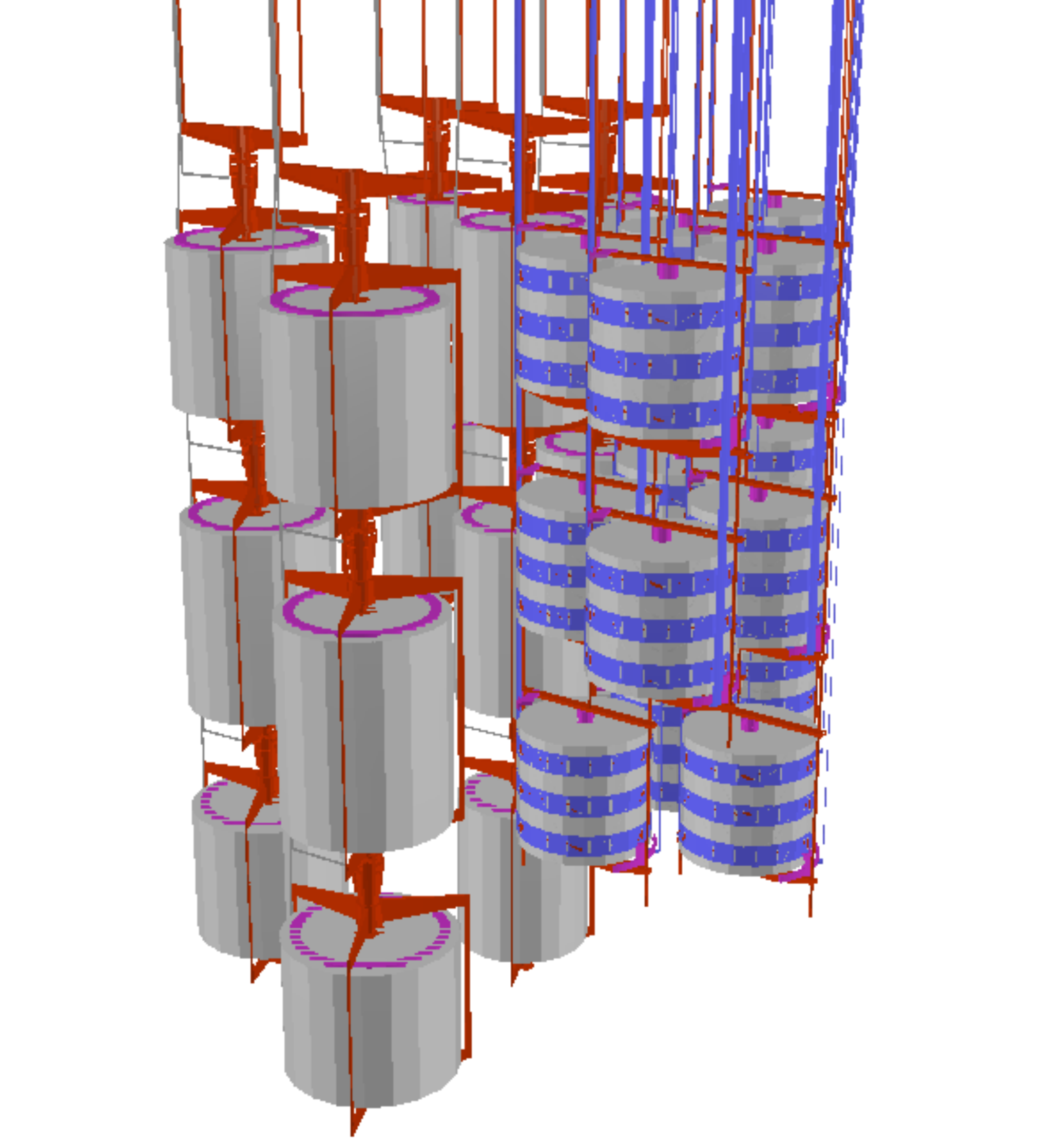}
\caption{Rendering of \Gerda\ detector array with cryostat and shields
removed. On the right hand side are 18-fold segmented HPGe detectors,
while left are unsegmented ones. Detectors of the same type are only
modeled once. The codes are reused to construct the whole array.}
\label{fig:GerdaArray}
\end{figure}

These are the physical geometries of the detectors or experiments that are being simulated. \MaGe\ 
supports geometric description via the \GF\ geometry description classes or through an 
interface with the Geometry Description Mark-up Language~\cite{gdml} (GDML). 
\MaGe\ currently has about 30 user-selectable geometries. The geometries 
range from simple cylindrical crystals to a full detector array with 60 crystals, mounting 
components, shield and surrounding room. Each geometry is encoded in a class that derives 
from a geometry base class that contains the following basic components of a
geometry:
\begin{itemize}
\item A unique identifying serial number string.
\item A detector name string. 
\item A \emph{ConstructDetector()} method that is invoked by \GF\ to construct the detector geometry during run-time. 
\item An associated \emph{G4LogicalVolume}.
\item A setting of the importance value of the region, used by \GF\ 
when performing simulations that require importance sampling of geometries to optimize performance. 
\end{itemize}
This design also allows the reuse of existing geometry classes, since
the classes describing a geometry can be instantiated within a class
that requires that component. For example, a detailed germanium
crystal has been coded that is used many times in other simulated
detector geometries. This crystal can be simulated on its own, or be
instantiated many times in a complex detector array. Shown in
Fig.~\ref{fig:MJDEMOGeometry} is an example of how multiple
stand-alone geometry classes are combined to create a complex detector. A rendering of the detector is shown in Fig.~\ref{fig:MJDEMOrendering}. The \Gerda\ detector array rendering shown
in Fig.~\ref{fig:GerdaArray} is constructed in a similar way. A bonus
of this approach is that the \Gerda\ and \MJ\ collaborations can share
the same basic geometries, such as crystals. This reduces redundant
code and increases code scrutiny.

\subsection{Event Generators}
\label{se:event_generators}
These classes generate the initial conditions of each event to be simulated. \MaGe\ has several event generators that all inherit from an \emph{MGVGenerator} base class. One such generator is instantiated by the user at run-time via a \GF\ messenger. The \emph{MGVGenerator} base class contains the following relevant virtual methods that are defined in the daughter class:
\begin{itemize}
\item A \emph{BeginOfRunAction()} method that is executed at the beginning of a \GF\ run. It is primarily used to compute lookup tables from analytic expressions for distributions to be sampled.
\item A \emph{EndOfRunAction()} method that is executed at the end of a run. It is primarily used to deallocate memory used by lookup tables. 
\item A \emph{GeneratePrimaryVertex(G4Event *event)} method that passes the initial event particle type, position and 
momentum to the \emph{G4Event} object. 
\item A \emph{SetParticlePosition(G4ThreeVector vec)} method that can be used by another generator, such as the position sampler described below, to set the position for the initial vertex. 
\end{itemize} 
The generator may also have a \GF\ messenger associated with it to allow the user to set parameters at run-time. Using this base class, we have created the following specific generators for \MaGe:
\begin{itemize}
\item A wrapper for the Radioactive Decay Module~\cite{G4RDM} 
(RDM) in \GF\ that generates radioactive decays from unstable nuclei at a point.
\item ``Hybrid" generators that use the RDM generator to generate the initial decay of a nucleus, but then uses a customized surface or volume sampler to place it at a specific location in our simulated geometry. We developed a volume sampler that can generate points uniformly distributed in any {\sc
Geant4} solid. This is required to
simulate radioactive contamination embedded in detector components. A
surface sampler was also developed that creates points uniformly
distributed on the surface of any {\sc Geant4} solid. 
This surface sampler is complex and is described further in~\cite{surfacesampler}. 
A surface sampler is required to simulate surface contaminations, in particular $\alpha$-emitters and $\beta$-emitters on detector surfaces inside the cryostat that are within line-of-sight of the HPGe crystals.
\item Generators to simulate neutron and
muon backgrounds in underground laboratories, using either theoretical
models or data-driven approaches. Interfaces are available which read
initial conditions for an event from other codes, such as
\textsc{Sources4A}~\cite{sources4a} for neutron flux and
\textsc{Musun}~\cite{musun} for muon flux, as well as other selected analytical models, such as that by Wang et al.~\cite{wang}. 
\item A wrapper for the FORTRAN-based \textsc{Decay0}~\cite{decay0} generator. It is primarily used to generate 
different types of double-beta decays, but can also simulate normal nuclear decays. \textsc{Decay0} is able to 
simulate properly the angular correlation between $\gamma$-rays in nuclear de-excitation cascades, that is not accounted for  
in the \GF\ Radioactive Decay Module.
\item A customized generator to simulate the neutron and gamma flux from an AmBe source used for studying neutron interactions. 
\end{itemize}

\subsection{Physics Lists}
\label{se:physics_lists}
A collection of {\sc Geant4} physics
processes is called a physics list. They define the particles that are
included in the simulation and the decays and interactions they can
undergo.  There are several physics lists implemented in {\sc MaGe},
each optimized for the particular problem being simulated.  One list
is optimized at higher energies for simulating cosmic-ray muon
interactions, while others are optimized for standard electromagnetic
interactions at lower energies, i.e. below 10 MeV. These are used to
simulate the response of detectors to the decay of radioactive
isotopes. Each physics list is contained in its own class that is
instantiated at runtime. The physics lists and requirements are discussed in more detail in section~\ref{section:physics}.
 
\subsection{Output Format}
\label{se:output_format}
During the simulation \GF\ generates complete information about the trajectory and interactions of particles as they propagate through the detector. Although all of this information is available to the user, it is typically processed, parsed and saved to an output file for further analysis after the simulation run is complete. Different detector or even different simulations of the same detectors have specific output requirements. \MaGe\ defines a generic base output class and 
does not have a built-in output format, but we have implemented interfaces to the AIDA-compliant~\cite{aida} 
and ROOT~\cite{root} analysis tools, as well as simple text-based output.  
%Each detector and Monte Carlo study has unique
%output requirements. {\sc MaGe} has several different types of output
%formats that can be combined to provide the information relevant to a
%specific study. 
At one extreme \MaGe\ has implemented a class that saves all of the \GF\ information for each step, at the other end is an output class that only generates a histogram of energy deposited in a single HPGe crystal. Each output format consists of a class that inherits from a virtual base class (\emph{MGVOutputManager}) that contains the following main methods:
\begin{itemize}
\item \emph{BeginOfRunAction()} and \emph{EndOfRunAction()} methods that are executed at the beginning and end of a simulation run respectively. They are used to open and close data files, create file formats, 
i.e. ROOT trees, create histograms, and allocate and deallocate data structures.
\item \emph{BeginOfEventAction()} and \emph{EndOfEventAction()} methods that are executed at the beginning and end of the events. They are used to clear arrays that store stepping information and to perform processing on event data and fill event histograms. 
\item A \emph{SteppingAction()} method that is executed at the end of each step. It gathers all the information concerning the particle in the given track and adds it to a histogram, or accumulates it in a variable. 
\end{itemize}
These classes, in turn, can be
inherited by classes that simulate and store detector responses and
save any relevant information. 

\subsection{Materials}
\label{se:materials}

\GF\ defines materials internally in terms of how they interact 
with ionizing radiation, but \MaGe{} adds information about radioactive contaminants and other properties 
of interest to \MJ\ and \Gerda.
{\sc MaGe} has the ability to read in all
relevant information about materials from a PostgreSQL database. This
is currently limited to quantities required by \GF\ such as density, isotopic
abundance, etc. Once the \textsc{Majorana} or \textsc{Gerda} 
detectors are constructed,
the materials used will be carefully assayed and
characterized. All this information will be saved in a database as
well. {\sc MaGe} can then use this information to include the measured
activities in the simulation on a component-by-component and time-dependent level,
reducing systematic uncertainties in sensitivity calculations. \\

\subsection{Example}
\label{se:simple_example}

\MaGe\ is typically compiled into a single executable. 
The user selects the particular instance of the components they require during run-times via macros saved as text files. 
\MaGe\ macros are based on the \GF\ \emph{messenger} classes. A simple, self-explanatory example of such a macro is given below. The simulation consists of a simple block that is bombarded with a beam of neutrons. The scattered neutrons are analyzed by the output class and elastic cross-sections are computed. This example is used to verify the neutron cross-sections implemented by \GF.

{\small
\begin{verbatim}
% Select geometry component
% and set parameters describing 
% its geometry.
/MG/geometry/detector solidBlock
/MG/geometry/solidblock/material Hydrogen
/MG/geometry/solidblock/edgeLength 1.0 cm

% Select output format component.
/MG/eventaction/rootschema HPNeutronTest

% Select generator component.
/MG/generator/select SPS

% Start run
/run/initialize
/run/beamOn 500000
\end{verbatim}
}

% ---------------------------------------------------------------------
% physics
% ---------------------------------------------------------------------

\section{Physics} \label{section:physics}

The physics list in {\sc MaGe} has been optimized for the reliable
simulation of the most common background
sources in $0\nu\beta\beta$-decay experiments. It was designed
according to the suggestions of the \textsc{{\sc Geant4}}
team~\cite{physics_list} and optimized for
low-background physics
applications~\cite{Bauer:2004an,Pandola:2007}.
The default  \MaGe\ physics list is mainly based on the
Underground Physics advanced example which is distributed with {\sc
Geant4}~\cite{physics_list}.

By default, \MaGe\ uses the Low-Energy models based on the Livermore data 
libraries~\cite{livermore1,livermore2} for the description of the electromagnetic 
interactions of electrons, $\gamma$-rays and ions. These models include atomic 
effects (e.g. fluorescence and Doppler broadening) and can handle interactions of electrons, positrons and photons 
with energies down to 250~eV. \\
For specialized applications, electromagnetic interactions of $\gamma$-rays, 
electrons and ions can be simulated in \MaGe\ by the so-called ``standard models'' provided 
by \GF~\cite{physics-manual}. These models are tuned to high-energy physics applications; 
they are less precise in the low energy region and do not include atomic effects. 
However, they are faster in terms of computing time.

\MaGe\ uses the standard \GF\ 
models for the electromagnetic interactions of muons and of positrons. 
Furthermore, synchrotron radiation is included in the physics list for electrons and positrons. 
The electromagnetic physics processes provided by \GF\ for $\gamma$-rays and e$^{\pm}$ 
(both "standard" and "low-energy") have been systematically validated by the 
\GF\ Collaboration~\cite{Amako:2005} and by other groups~\cite{Poon:2005} at the 
few-percent level. The precision of the electromagnetic models for muons is discussed 
in ~\cite{Bogdavov:2006}.

Interactions of e$^{\pm}$ and $\gamma$-rays with nuclei are simulated on the basis of the 
equivalent photon approximation. The reaction cross section is determined according 
to a parameterization from experimental data for all incident energies from the hadron 
production thresholds upwards. Different energy regimes are considered separately 
in the parameterization, as described in Ref.~\cite{physics-manual}. Hadronic final 
states are generated using a Chiral Invariant Phase Space (CHIPS) decay 
model~\cite{chips3}.

For energies above 3.5~GeV, the final states of photo-nuclear reactions are generated 
according to a theory-based parton-string model, called the Quark-Gluon String Precompound 
model (QGSP). The model is composed of several components: the quark-gluon string 
(QGS) part handles the formation of the initial strings in the initial collision~\cite{qgsp1}; 
string fragmentation into hadrons is handled by the quark-gluon string fragmentation 
model~\cite{qgsp2}, while the pre-compound model~\cite{qgsp3} takes care of 
the de-excitation of the residual nucleus.

Hadronic interaction of muons with nuclei are managed by the \texttt{G4MuNuclearInteraction} 
model. Muons produce virtual photons which are in turn converted to pions which interact with 
the nucleus using a model derived from 
the \textsc{Gheisha} code~\cite{gheisha}. The physics list also includes 
the capture process of $\mu^{-}$ by nuclei. The hadronic interactions included in the \MaGe\ physics list handle elastic scattering, 
inelastic scattering, capture (for neutrons, $\pi^{-}$ and K$^{-}$), fission (neutrons only) and 
decay.

The elastic scattering of all long-lived hadrons is described by the \texttt{G4LElastic} model, 
which is based on the \textsc{Gheisha} code~\cite{gheisha} and includes a 
parameterization for cross section and final state. 
Elastic scattering of neutrons from thermal energies to 20~MeV is simulated 
according to the data-driven \texttt{G4NeutronHPElastic} model, which is based on the tabulated 
cross section and final state data from the ENDF/B-VI database~\cite{endf1, endf2}.

Different energy regimes have been 
considered in the \MaGe\ physics list to simulate the inelastic interactions of long-lived 
hadrons. Each energy regime has its own specialized model. 
In particular:
\begin{enumerate}
\item theory-driven quark-gluon string precompound model (QGSP) for pions, kaons 
and nucleons in the high-energy region, up to 100~TeV.
\item Low-energy parametrized (LEP) model for pions and nucleons with energies between 10 and 
12~GeV and for kaons below 25~GeV. Such a model is applicable below about 30~GeV and is 
derived from the \textsc{Gheisha} package~\cite{gheisha}. The cross section and the final state are 
determined by parametrized functions which are fitted to experimental data.
\item Bertini (BERT) cascade model, based on a re-engineering of the 
\textsc{INUCL} code~\cite{inucl}, to describe nucleon and $\pi$ interactions below energies of 10~GeV.
The model includes the Bertini intra-nuclear cascade model with 
excitons~\cite{bert1, bert2}, a pre-equilibrium model~\cite{gri1,gri2}, a nucleus explosion 
model~\cite{breakup} and an evaporation model~\cite{weiss}. 
\item data-driven high-precision (HP) model for neutrons from thermal energies up to 20~MeV, 
based on the ENDF/B-VI database.
\end{enumerate}
Such an ensemble of models for hadronic inelastic interactions is shortly labeled as 
QGSP\_LEP\_BERT\_HP.\\

Alternative hadronic physics lists, which differ in the models used to describe the 
inelastic interactions of nucleons, are available in \MaGe\ and can be instantiated by 
messenger commands. They are: 
\begin{itemize}
\item QGSP\_LEP\_BIC\_HP, which employs the Binary cascade~\cite{physics-manual} model  
(instead of the Bertini cascade model) for inelastic interactions of nucleons below 
10~GeV. The model handles the intra-nuclear cascade as well as the remaining 
fragment, which is treated by precompound and de-excitation models. Cross sections are 
parametrized using experimental data. This physics list is similar to the one used in 
Ref.~\cite{Lindote:09} for the simulation of muon-induced neutrons, the only difference 
being that the list of Ref.~\cite{Lindote:09} does not include the LEP bridge between 
the high-energy QGSP regime and the low-energy Binary cascade.
\item QGSP\_LEP\_HP, where parametrized LEP models are used for nucleons below 10~GeV, instead 
of intra-nuclear cascade models. Such a list is much faster than the default one from 
the point of view of the computation time.
\item QGSC\_BERT\_HP, which employs the quark-gluon string CHIPS model (QGSC) instead of 
the quark-gluon string precompound model (QGSP) to simulate high-energy inelastic 
interaction of nucleons (from 10~GeV to 100~TeV). The QGSC model differs by the QGSP 
by the fact that the de-excitation of the residual nucleus is handled by a CHIPS 
model~\cite{chips3,chips1,chips2}, 
rather than the pre-compound model. The physics list does not include 
the LEP parametrized model. The QGSC model is also used to produce the final state 
following photo-nuclear interactions above 3.5~GeV.
\end{itemize}

\begin{figure*}[tbh]
\centering
\includegraphics[width=0.70\textwidth]{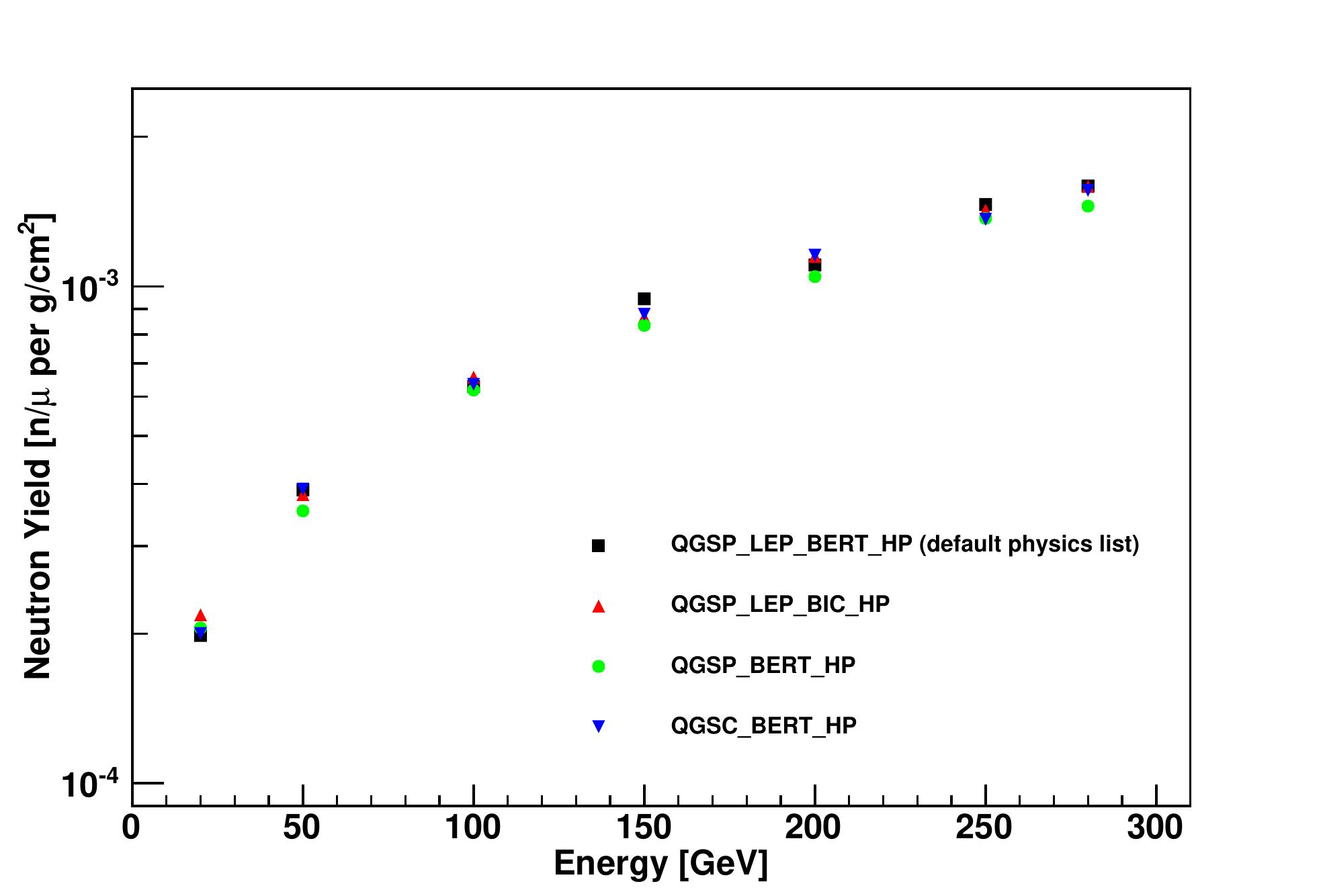}
\caption{Neutron yield from muon-induced showers in metallic germanium.
\MaGe\ had been run with the version 9.0 of \GF.}\label{fig:germanium_mc} 
\end{figure*}

Multiple alternative lists for hadronic physics based on independent models are used 
in \MaGe\ mainly for testing purposes, namely to cross-check results and evaluate 
systematic uncertainties from the simulations. Crucial background sources for many 
underground experiments are due to high-energy interactions of cosmic-ray muons.
Background estimates (e.g. production of secondary neutrons or long-lived 
unstable nuclei) often rely on the ability of simulation codes to model high-energy 
hadronic and electromagnetic showers initiated by muon interactions. Fig.~\ref{fig:germanium_mc} shows 
the neutron production yield by muons in Germanium 
vs. the muon energy for the four hadronic physics lists provided by \MaGe\ (one default and 
three alternative). The simulations are run with \GF\ 9.0. 
Simulation results are very similar (within 15\%) in the full energy range, 
indicating that details in the physics modeling of high-energy hadronic interactions do not have a large 
impact on some parameters of interest for underground physics experiments. 
Neutron yield in Ge obtained with \MaGe\ (about $1.5\cdot10^{-3}$ neutrons per muon 
per g/cm$^{2}$ at 280~GeV) 
%is larger than the value derived from Fig.~5 of Ref.~\cite{Araujo:05} 
%(about $1.1\cdot10^{-3}$ neutrons per muon per g/cm$^{2}$), which was obtained with an older 
%version of \GF. The value derived in this work 
is consistent ($< 10\%$) with the results based on the \textsc{Fluka-1999} 
code~\cite{fluka}, derived by the power law parameterization of Fig.~5 of 
Ref.~\cite{Araujo:05}. \\

The \Gerda\ experiment uses a water Cherenkov muon veto and internal 
shield consisting of liquid Ar. For this reason, \MaGe\ takes advantage of 
\GF\ ability to simulate optical
photons. While the default {\sc MaGe} physics list does not include
interactions of optical photons these processes can be enabled during
runtime. The underlying models encompass scintillation light emission
(possibly with different light yields for electrons,
$\alpha$-particles and nuclei), Cherenkov light emission, absorption,
boundary processes, Rayleigh scattering and wavelength shifting. 
If optical photon treatment is enabled, it is
necessary to specify all relevant optical properties of interfaces and
bulk materials (refraction index, absorption length, etc.) in the
geometry definition. \\

{\sc Geant4} tracks all simulated particles down to zero range, although
various options exist to manually limit step size, track length, time-of-flight, 
and other parameters. Production cuts for
$\delta$-rays and for soft bremsstrahlung photons are
expressed in spatial ranges and are internally converted into energy
thresholds for the production of soft photons and $\delta$-rays in the
corresponding material. It is necessary to find a trade-off between
accuracy and computing time in most applications. Therefore {\sc MaGe}
provides three production cut realms: DarkMatter, DoubleBeta and
CosmicRays. The DarkMatter realm is used for high-precision
simulations, especially related to background studies for dark matter
applications and surface effects: the cuts for $\gamma$-rays and $e^{\pm}$  
are 5~$\mu$m and 0.5~$\mu$m, respectively, corresponding to
a $\sim$1~keV energy threshold in metallic germanium. The DoubleBeta realm
({\sc MaGe} default) is suitable for signal and background studies
related to double-beta decay, i.e. in the MeV energy-region: the
range cut for $\delta$-ray production is relaxed to 0.1~mm,
corresponding to a 100~keV threshold in metallic germanium. The
CosmicRays realm is used for the simulation of extensive
electromagnetic showers induced by cosmic-ray muons. The
cut-per-region approach is used in this setup. Sensitive regions are
defined for which the production cuts are the same as for the
DoubleBeta realm. They are more relaxed everywhere else (5~cm for
$\gamma$-rays and 1~cm for $e^{\pm}$). By avoiding the
precise tracking of particles in the inactive detector components,
computing time is saved. \\

     {\sc MaGe} includes some provisions to improve agreement between 
simulation and experimental results.  
For instance, simulations do not account 
for inefficient conversion of germanium nuclei recoil energy to 
ionization energy, also known as quenching.  {\sc MaGe} contains output classes that simulate this conversion inefficiency using the parameterizations from~\cite{ljungvall}. \\
 
{\sc Geant4} performs simulations on an event-by-event basis, where
each event begins with the release of a particle from a generator, and
ends when the interactions of the primary particle and its secondaries
have finished.  When long-lived radioactive decays occur, a single
{\sc Geant4} event may span many simulated years.  Output classes in
{\sc MaGe} divide {\sc Geant4} events into intervals that span user-selectable
times.  The total energy deposited during specific time intervals can
be reported.  This information can be used to simulate the
effectiveness of timing cuts at removing backgrounds.  It can also
improve agreement between results of {\sc MaGe} simulations and
experimental data.  Simulated energy deposits occurring long after the
duration of an experiment can be excluded.  Simulated energy deposits
in close succession can be summed to approximate the pile-up due to the finite time resolution
of data acquisition hardware used in an experiment. %\\

% ---------------------------------------------------------------------
% validation 
% ---------------------------------------------------------------------

\section{Validation of the simulation}
\label{section:validation} 

The {\sc Geant4} simulation toolkit is used in various applications of
modern physics. These range from simulations in high-energy particle
physics to astrophysics and medical science. In parallel to the
development of new simulation modules the verification of the
simulation code is an important task for developers and users. Several
modules have been developed to describe the interactions of low energy
photons, electrons and hadrons with matter. These
are of particular importance for applications such as {\sc MaGe} and
are tested within the two collaborations developing the software. In
the following, the current status of {\sc MaGe} validation efforts is
summarized. \\

\begin{figure}[tbh]
\centering
\includegraphics[height=0.48\textwidth,angle=90]{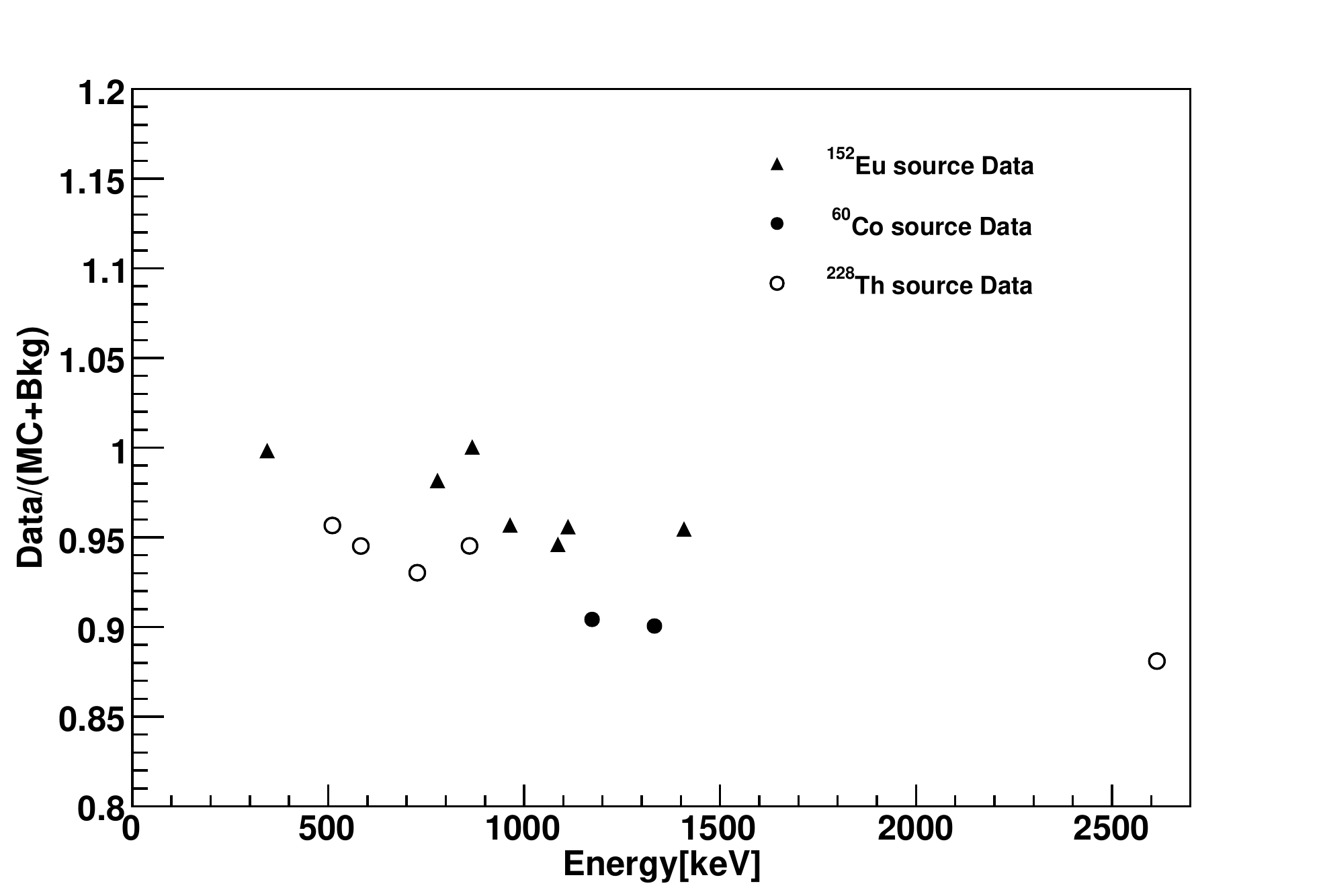}
\caption{Difference between number of events in characteristic photon peaks from $\rm ^{60}Co$, $\rm ^{152}Eu$ and $\rm ^{228}Th$ sources plus background and a simple~\MaGe~simulation. The maximal deviation was found to be approx. 12\%.}
\label{fig:cryoliquidsubmersion_mc} 
\end{figure}

The description of electromagnetic interactions in the energy region
up to several MeV was tested with high-purity germanium detector
systems. A reverse electrode coaxial germanium detector was operated
directly submerged in liquid nitrogen in the so-called cryoliquid-submersion
test stand~\cite{KKPHD}.  The crystal was exposed to the radioactive
sources $^{60}$Co, $^{228}$Th and $^{152}$Eu.  The comparison between
a simple \MaGe~simulation and recorded data is shown in
Fig.\ref{fig:cryoliquidsubmersion_mc}. The biggest deviation was found
to be approximately 12\%. Most probably the decrease of
$Data/(MC+Bkg)$ with increasing energy is due to the fact that the
inactive material between source and detector, e.g. dewar walls, is
not precisely known.\\

The most extensive verification effort was performed with an 18-fold
segmented \textsc{Gerda} prototype detector. The segmented germanium
crystal was operated in vacuum and exposed to several radioactive
sources ($^{60}$Co, $^{228}$Th,
$^{152}$Eu)~\cite{Abt:2007rg,Abt:2007rf}.  A large fraction of the
emitted $\gamma$-rays deposit energy in more than one segment. This
feature allows these events to be distinguished from those which
deposit energy in relatively small volumes, such as
\nubb\ decays.  Such segmentation-based discrimination
between single- and multiple-site interactions was compared between
experiment and simulation, with deviations found on the
5\%~level. \\

\MaGe\ has been used to simulate the response of a variety of low-background assay detectors in use by the \MJ\ collaboration. In general the simulation agreed with the data to a few percent and the largest discrepancies were ascribed to uncertainties in the geometry models that have been coded in the 
simulation. 
The detailed shape of the Compton continuum for a HPGe detector exposed to variety of sources was also studied in~\cite{evans}. The simulation and data compared favorably in this study.\\ 

The description of neutron interactions with Ge is probed by comparing data
from a measurement of an AmBe source with predictions from {\sc
MaGe}. The measurements have been performed with a {\sc Clover}
detector and with the 18-fold segmented detector cited 
previously~\cite{siegfried_neutron}. At an energy of several
MeV, neutrons mostly interact through elastic and inelastic scattering
as well as neutron absorption. The measured energy spectra were
studied and photon lines from neutron interactions with the germanium
detector itself and the surrounding materials were
identified~\cite{Mei:2007zd}. 
A few discrepancies in the \GF\ simulation were identified:
\begin{itemize}
\item the 2223.0-keV peak from H(n,$\gamma$)D appears at 2224.6 keV 
in {\sc Geant4} simulations (bug report \# 955)~\cite{G4bug};
\item meta-stable nuclear states are not produced by {\sc Geant4} 
as a result of neutron interactions (bug report \# 956)~\cite{G4bug};
\item neutrons do not produce internal conversion electrons in 
{\sc Geant4} (bug report \# 957)~\cite{G4bug}.  
\end{itemize}
The first and third bugs are related to issues in the G4NDL nuclear data files distributed with
\GF, and can be solved by editing or augmenting the data library. 
The third bug also involved several coding errors, corrections for which were provided by
the \MaGe\ group. Fixing the second bug requires functionality that the \GF\ Collaboration does
not plan to implement at this time. 
Nevertheless, a viable solution has been developed by the \MaGe\ 
group. The work-around for the meta-stable problem consists in identifying 
at run-time those neutron interactions that might produce a meta-stable nucleus. In this 
case, the new nucleus track is discarded, since \GF\ would always generate it in 
the ground state, but its position and the parent neutron energy are logged in a file. 
Later on, a new simulation job is launched which uses the position information from 
the previous one and where the ratio between ground and meta-stable nuclear states 
is set manually to the proper value.  \\

{\sc MaGe} was also used to study and verify the
simulation of spallation neutron production and propagation. At the
CERN NA55 experiment the neutron production from a 190~GeV muon beam
incident on different targets was measured. At the SLAC electron beam
dump experiment the neutron propagation through different thicknesses of concrete was
measured. Both experiments were simulated within the {\sc MaGe}
framework. The attenuation of the neutron propagation was found to be larger in the simulation 
than measured in the SLAC experiment~\cite{mike_neutron}. A method to correct the neutron 
over-attenuation in \textsc{MaGe}-based simulations was implemented in \MaGe\ and is described 
in~\cite{mike_neutron}. \\

\begin{figure*}[tbh]
\centering
\includegraphics[width=0.70\textwidth]{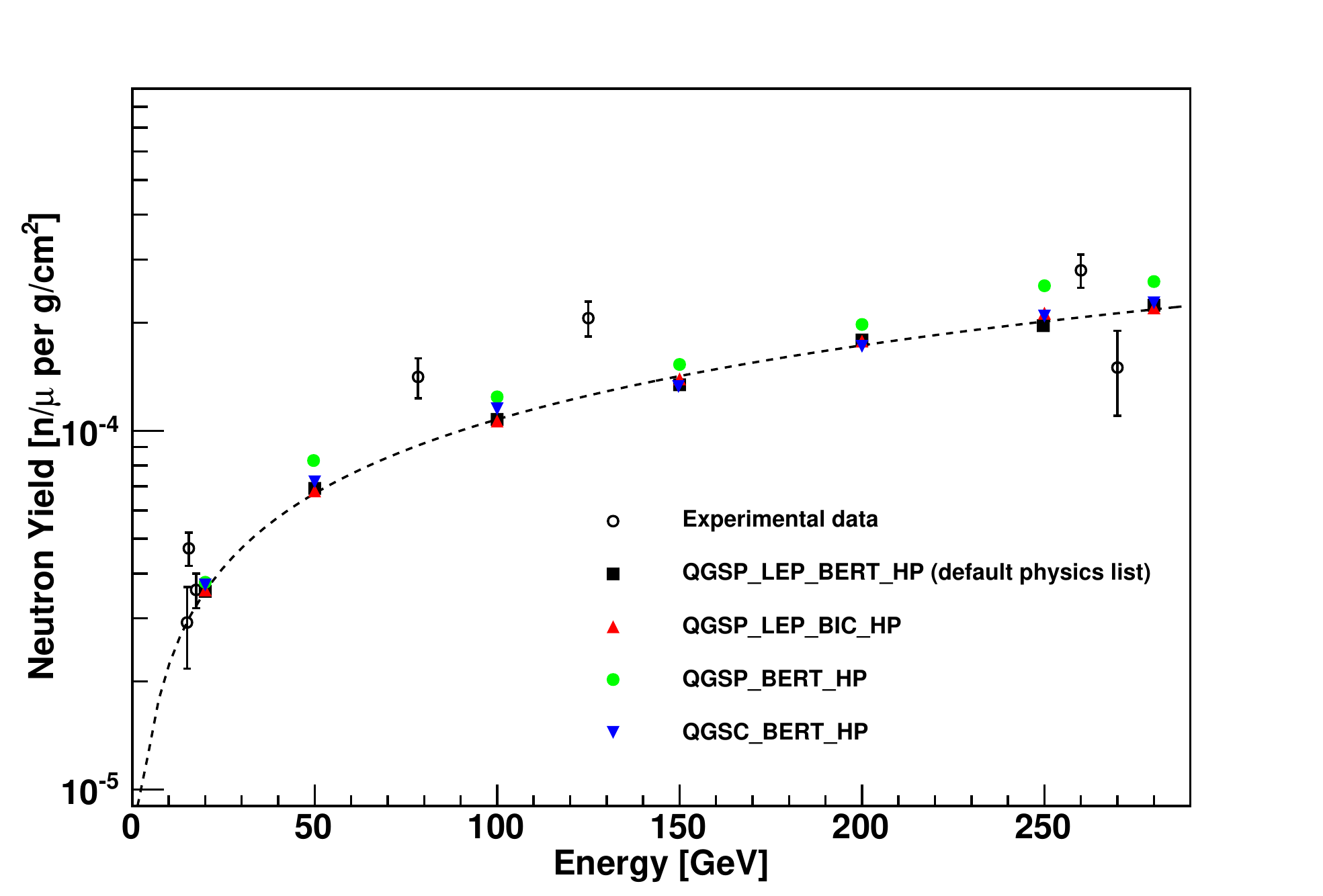}
\caption{Neutron yield from muon-induced showers in liquid scintillator (C$_{10}$H$_{22}$). 
Experimental data are from the Refs.~\cite{mudata} (reported 
in~\cite{Araujo:05}) and from Ref.~\cite{kamlanddata}. 
Monte Carlo results 
(default physics list) are well parametrized by a scaling law $E^{0.685}$ 
(dashed curve). \GF\ 9.0 had been used for this study.}\label{fig:scintillator_vs_data} 
\end{figure*}

It was also found that \textsc{MaGe}/\textsc{Geant4} underestimates 
the neutron production from muon interactions measured by NA55, especially 
in high-Z materials, by more than a factor of two~\cite{mike_neutron}. 
Results obtained in the \textsc{MaGe} simulation of NA55 have been 
compared with the \textsc{Geant4}- and \textsc{Fluka}-based~\cite{fluka} 
Monte Carlo simulations of the same experiment performed in~\cite{Araujo:05}, and 
found to be consistent.
The disagreement between Monte Carlo simulation codes and NA55 data for muon-induced 
neutron production is discussed in detail in~\cite{Araujo:05} and \cite{ilias06}.
The muon-induced neutron yield in lead has been recently 
re-measured at the Boulby underground laboratory~\cite{Araujo:08}. In this case,
in which the neutron energies are much lower, the 
\GF-based simulation is found to over-estimate the experimental neutron yield, 
which is the opposite behavior with respect to the NA55 data.
\MaGe\ results for muon-induced neutron production by muons in liquid scintillator 
have been compared to the set of experimental data reported in Ref.~\cite{Araujo:05}. 
Fig.~\ref{fig:scintillator_vs_data} displays the neutron production yield by muons derived 
by the \MaGe\ simulations 
(using the alternative hadronic physics lists described in Sect.~\ref{section:physics}) 
vs. energy, superimposed with the experimental data~\cite{mudata} reported in~\cite{Araujo:05} 
and the recent data from the KamLAND experiment~\cite{kamlanddata}. 
The agreement between \MaGe\ simulations and experimental data is typically 
better than a factor of two. Results are consistent with those obtained in the 
recent work Ref.~\cite{Lindote:09}. \\

%

% ---------------------------------------------------------------------
% conclusions 
% ---------------------------------------------------------------------

\section{Conclusions}
\label{section:conclusions}

We presented the {\sc MaGe} framework for simulating interactions in
neutrinoless double-beta decay experiments that utilize enriched HPGe
detectors. The benefits of {\sc MaGe} can be summarized as:
\begin{itemize}
\item Reliable Monte Carlo framework based on {\sc Geant4} for
low-background, low-energy experiments.
\item Ongoing tests of the code and validation of the physics processes. 
\item Flexible geometry and physics code framework that emphasizes code
reuse and verification. 
\item General purpose tools like surface 
and volume sampling, custom isotope decay generators, etc.
\end{itemize} 

In general there is good agreement between the \GF\ simulation with the MaGe physics list
and the measurements of electromagnetic interactions with average
discrepancies of the order of (5-10)\%. Several problems have been
identified in the simulation of neutron interactions.
These problems have been reported to the {\sc Geant4} collaboration
and are under investigation by the \MaGe\ developers.
\\

We anticipate that \MaGe\ will form the foundation of the
simulation and analysis framework of the \Gerda\ and \MJ\
experiments. This type of framework and its associated event generators and physics lists is also useful for other
low-background underground experiments, such as solar, 
reactor and geological neutrino experiments, direct dark matter searches, 
and other neutrinoless double-beta decay search.
These experiments share many detection techniques and background
issues in common with \textsc{Gerda} and \textsc{Majorana}.

% ---------------------------------------------------------------------
% Acknowledgment
% ---------------------------------------------------------------------

\section*{Acknowledgments}

The authors would like to express their gratitude to all members of the 
\textsc{Majorana} and \textsc{Gerda} collaborations, for their continuous 
feedback and useful advices.
The authors would like to thank V.~Tretyak and C.~Tull for useful
discussions, and I.~Abt for useful
discussions and her efforts to help unify the Monte Carlo of both
collaborations. This work has been supported by the
\textsc{ILIAS} integrating activity (Contract No.RII3-CT-2004-506222)
as part of the EU FP6 programme, by BMBF under grant 05CD5VT1/8, 
and by DFG under grant GRK 683.
This work was also supported by
Los Alamos National Laboratory's Laboratory-Directed Research and Development Program,
and by the Office of Science of the US Department of Energy 
at the University of Washington under Contract No. DE-FG02-97ER41020, 
at Lawrence Berkeley National Laboratory under Contract No. DE-AC02-05CH11231, at the University of North Carolina under Contract No. DE-FG02-97ER41041, 
and at Pacific Northwest National Laboratory under Contract No. DE-AC06-76RLO1830. 
This research used the Parallel Distributed Systems Facility at the 
National Energy Research Scientific Computing Center, which is 
supported by the Office of Science of the U.S. Department of Energy 
under Contract No. DE-AC02-05CH11231.

\ifCLASSOPTIONcaptionsoff
  \newpage
\fi

\end{document}